\documentclass[prd,preprint,showpacs]{revtex4-1}
\usepackage{amsmath}
\usepackage{latexsym}
\usepackage{amsfonts}
\usepackage{graphicx}
\usepackage{mathrsfs}
\usepackage{CJK}
\usepackage{hyperref}

\newcommand{\ud}{\mathrm{d}}
\newcommand{\pd}{\partial}
\newcommand{\xu}{\phantom}
\newcommand{\scri}{\mathscr I}
\newcommand{\scrih}{\Delta}
\newcommand{\ghor}{\Gamma^{ab}_{\xu{ab}cd}}
\newcommand{\lie}{\mathscr L}
\newcommand{\scrin}{\mathscr N}

\newcommand{\frakt}{\mathfrak t}
\newcommand{\frakb}{\mathfrak{bms}}
\newcommand{\frakl}{\mathfrak l}

\hypersetup{
  colorlinks = true,
  urlcolor = blue,
  linkcolor= blue,
  citecolor= green,
  filecolor= magenta,
}

\begin{document}
\begin{CJK*}{GB}{gbsn}

\title{Asymptotic Symmetries of the Null Infinity and the Isolated Horizon}

\author{Shaoqi Hou}
\email{shou1397@hust.edu.cn}
\affiliation{School of Physics, Huazhong University of Science and Technology,
Wuhan, Hubei 430074, China}

\begin{abstract}
The common intrinsic geometry shared by all the null hypersurfaces gives rise to the asymptotic symmetries found on the null infinities $\scri^\pm$ and the isolated horizons $\scrih$. In this work, the properties of a null hypersurface are reviewed and the invariance of its intrinsic geometry ($n^an^bh_{ab}$) is revealed under the spacetime conformal transformation. The generators, i.e., infinitesimal symmetries, of the conformal transformations tangent to the null hypersurface are defined and classified by their effects on the induced metric and the normal vector field. Two particular examples and their symmetries are discussed, that is, the null infinities $\scri^\pm$ of an asymptotic flat spacetime, and the isolated horizon $\scrih$ of a black hole.
\end{abstract}

\maketitle
\end{CJK*}

\section{Introduction}

There has been growing interest in identifying the ``asymptotic symmetries" on black hole horizons $\scrih$. These symmetries are similar to the asymptotic symmetries defined on the null infinities $\scri^\pm$ discovered by Bondi, van der Burg, Metzner \cite{Bondi21} and Sachs \cite{Sachs103,RN49} in 1960's, and dubbed as the Bondi-Metzner-Sachs (BMS) symmetries. In particular, Hawking, Perry and Strominger \cite{RN57} proposed that the supertranslations on the black hole horizon might be responsible for the black hole entropy and solve the information paradox \cite{RN810}. Several works \cite{RN397,RN946,RN951,RN401,RN952,RN945,RN192,RN398,RN1107,RN1108,RN1109,Afshar:2016uax,RN80} have also been done to look for the asymptotic symmetries on the black hole horizon using perturbative methods. In Ref.'s\cite{RN1047,RN1045}, the universal structure of the asymptotic Killing horizon is identified and used to formulate the asymptotic symmetries from a geometric point of view.

To certain extent, the very existence of the similar symmetries on $\scrih$ and $\scri^\pm$ is due to the intrinsic geometry shared by all null hypersurfaces. This work will attempt to reveal the common intrinsic geometry that leads to the similar symmetries on $\scri^\pm$ and $\scrih$. The null hypersurface is different from the non-null ones (timelike and spacelike) mainly in two aspects: 1) Its induced metric $h_{ab}$ is degenerate, so it is not sufficient to define its intrinsic geometry using merely $h_{ab}$, while the induced metric of a non-null hypersurface is non-degenerate, so its intrinsic geometry can be completely characterized by its induced metric. 2) The normal vector $n^a$ of a null hypersurface is also tangent to itself, but that of a non-null hypersurface does not belong to the tangent space of the non-null hypersurface. Therefore, $n^a$ also reflects the intrinsic geometry of the null hypersurface. In contrast, the normal vector to a non-null hypersurface describes its extrinsic geometry, i.e., how the hypersurface is embedded in an ambient manifold.

Although the degeneracy of the induced metric of a null hypersurface seems to make it impossible to access its intrinsic geometry, the combination of the induced metric and the normal vector $\ghor=n^an^bh_{cd}$ \footnote{First identified by Geroch in \cite{Geroch1977}. Defined more precisely later.} fully grasps the geometry of the null hypersurface. Moreover, this object possesses more symmetries than the spacetime metric $g_{ab}$, even if there are no spacetime symmetries at all, because it remains invariant under the conformal transformation,
\begin{equation}
  h'_{ab}=\omega^2 h_{ab},\quad  n'^a=\omega^{-1}n^a.
\end{equation}
These symmetries form an infinite dimensional group, which is called $\mathcal{BMS}$ group  for the null infinities $\scri^\pm$ \cite{Bondi21,Sachs103,RN49}, and the horizon $\mathcal{BMS}$ group for the isolated horizons $\scrih$.

In fact, the conformal symmetry is not the essential ingredient for General Relativity (GR) and some alternatives to GR, e.g., Horndeski gravity \cite{RN171} and Einstein-\ae ther theory \cite{RN1096}. It is a symmetry of a ``superspace" which is a collection of spacetimes $(M,g_{ab})$ equipped with null hypersurfaces $(\scrin,\ghor)$. For any pair of spacetimes $(M_1,g^{(1)}_{ab})$ and $(M_2,g^{(2)}_{ab})$, there exists a diffeomorphism $\psi$ such that $\psi(\scrin_1)=\scrin_2$ and the pushforward $\psi_*\Gamma^{(1)ab}_{\xu{(1)ab}cd}=\Gamma^{(2)ab}_{\xu{(1)ab}cd}$ with $\scrin_1\subset M_1$ ($\scrin_2\subset M_2$) and  $\Gamma^{(1)ab}_{\xu{(1)ab}cd}$ ($\Gamma^{(2)ab}_{\xu{(2)ab}cd}$) the invariant structure for $\scrin_1$ ($\scrin_2$). In this sense, $\psi$ is a symmetry on $\scrin$.



In the present work, the intrinsic geometry of an arbitrary null hypersurface is identified and its invariance under the conformal transformation defines the asymptotic symmetries of the null hypersurface. The generators of the asymptotic symmetries form an Lie algebra donated as $\frakb$, which contains an infinite dimensional Abelian ideal $\frakt$ consisting of supertranslations. Two special examples of the null hypersurface are considered, i.e., the null infinities $\scri^\pm$ and the isolated horizons $\scrih$. For these examples, the quotient algebra $\frakl=\frakb/\frakt$ is isomorphic to the 6-dimensional Lorentz algebra. $\frakb$ thus resembles the Poincar\'e algebra in the flat spacetime and the conserved charges associated with these symmetries can be meaningfully defined.

This work is organized as follows: Section \ref{secgen} will be devoted to the discussion of the general properties and the symmetries of a generic null hypersurface $\scrin$ by reviewing and generalizing Geroch's treatment of the asymptotic symmetries on $\scri$ in Ref.\cite{Geroch1977}. Sections \ref{secnull} and \ref{sechor} discuss the special properties of the null infinities $\scri^\pm$ and the isolated horizon $\scrih$ to explicitly construct the infinitesimal symmetries, respectively.

For simplicity, the null infinities $\scri^\pm$ will be donated as $\scri$, for $\scri^-$ shares the similar structure with $\scri^+$.  Penrose's abstract index notation \cite{Penrose1984v1} is used in this work.

\section{The Null Hypersurface and the Symmetries}\label{secgen}

Both the isolated horizons $\scrih$ and the null infinities $\scri$ share some similarities, so it is appropriate to study the structures and symmetries of a generic null hypersurface $\scrin$. This section reviews and extends Geroch's approach \cite{Geroch1977}. In his original treatment \cite{Geroch1977}, the null hypersurface $\scrin$ is $\scri$. However, in the present work, $\scrin$ is a general one, and the only essential ingredients for the existence of the asymptotic symmetries are discussed.

\subsection{Basic setup}

Let $(M,g_{ab})$ be a spacetime, and an embedding $\phi:M\rightarrow\tilde{M}$ induces the conformal transformation,
\begin{equation}\label{confdef}
\tilde{g}_{ab}=\Omega^2\phi_*g_{ab},
\end{equation}
where $\phi_*$ is the pushforward induced by $\phi$, $\Omega:\tilde{M}\rightarrow\mathbb R$ is a smooth function and called the conformal factor. The manifold $(\tilde M,\tilde g_{ab})$ is called the unphysical spacetime. A further conformal transformation can also be made, and will bring $\tilde g_{ab}$ to
\begin{equation}\label{gag}
  \tilde g'_{ab}=\omega^2\tilde{g}_{ab},
\end{equation}
with $\omega$ a smooth, nonvanishing function on $\tilde M$. This reflects the gauge freedom in choosing a conformal factor. Let the covector field $n_a$ be the normal covector to $\scrin$. There is also a gauge freedom in choosing the normal covector as $n'_a=\omega n_a$ is also normal to $\scrin$.

In order to emphasize the important role of the intrinsic geometry of $\scrin$, it is convenient to consider $\scrin$ as the image of an inclusion $i$ of some 3-dimensional manifold $N$ into $\tilde M$. The intrinsic geometry of $\scrin$ is completely captured by $N$ via the pullback map $i^*$. However, not all the tensor fields on $\tilde M$ can be pulled back by $i^*$ to $N$. The functions $f:\tilde M\rightarrow\mathbb{R}$ and ($0,l$)-type tensors on $\tilde M$ can be pulled back, as $i^*$ is equivalent to restriction, e.g., $i^*f=f\circ i$, and $i^*n_a=0$. If a vector field $\tilde v^a$ is tangent to $\scrin$, i.e, $\tilde v^an_a=0$, it can be pulled back, and the result is denoted by $\hat v^a=i^*\tilde v^a$. In particular, $\tilde n^a=\tilde g^{ab}n_b$ can be pulled back: $\hat n^a=i^*\tilde n^a$. If a generic ($k,l$)-type tensor $\tilde T^{a_1\dots a_k}_{b_1\dots b_l}$ satisfies the property that for any $i^*(\tilde\nu_{a_1\dots a_k})=0$, $i^*(\tilde T^{a_1\dots a_k}_{b_1\dots b_l}\tilde\nu_{a_1\dots a_k})=0$ holds, then its pullback is defined according to
\begin{equation}
  i^*(\tilde T^{a_1\dots a_k}_{b_1\dots b_l}\tilde\mu_{a_1\dots a_k})=i^*(\tilde T^{a_1\dots a_k}_{b_1\dots b_l})i^*(\tilde\mu_{a_1\dots a_k}),
\end{equation}
where $\tilde\mu_{a_1\dots a_k}$ is also arbitrary. The pullback $i^*$ commutes with the Lie derivative, so $i^*\lie_{\tilde v}\tilde\mu_{a_1\dots a_l}=\lie_{\hat v}i^*\hat\mu_{a_1\dots a_l}$ provided that $\tilde v^a$ is tangent to $\scrin$. The pullback also commutes with the exterior derivative $\ud_a$. In the following, the pullback of a tensor field will be hatted.

Let $\hat h_{ab}$ be the pullback of $\tilde g_{ab}$, i.e., $\hat h_{ab}=i^*\tilde g_{ab}$. $\hat h_{ab}$ can be viewed as the induced metric on $N$ and is degenerate, as $\hat h_{ab}\hat n^b=i^*(\tilde g_{ab}\tilde n^b)=0$. Define $\hat h^{ab}$ to be the ``inverse" of $\hat h_{ab}$ in the sense that
 \begin{equation}\label{ingdef}
  \hat h_{ac}\hat h^{cd}\hat h_{db}=\hat h_{ab}.
 \end{equation}
So $\hat h^{ab}$ is unique up to an addition of $2\hat v^{(a}\hat n^{b)}$ with $\hat v^an_a=0$. In fact, the geometry of $N$ is characterized by $\hat n^a$ and $\hat h_{ab}$, and the combination $\ghor=\hat n^a\hat n^b\hat h_{cd}$ is an invariant tensor under the conformal transformation,
\begin{equation}\label{conft}
  \hat h'_{ab}=\omega^2\hat h_{ab},\quad \hat n'^a=\omega^{-1}\hat n^a.
\end{equation}
The symmetry of $N$ is defined as a diffeomorphism $\psi:N\rightarrow N$ such that $\psi^*\ghor=\ghor$. If a vector field $\hat\xi^a$ on $N$ generates such kind of diffeomorphism, it is called an \emph{infinitesimal symmetry} and satisfies,
\begin{equation}\label{defifsy}
  \lie_{\hat\xi}\ghor=0.
\end{equation}
All diffeomorphisms preserving $\ghor$ form a group donated as $\mathcal{BMS}$, and its Lie algebra $\frakb$ consists of the corresponding infinitesimal symmetries (e.g., $\hat\xi_1^a,\,\hat\xi_2^a$) with the Lie bracket given by $\lie_{\hat\xi_1}\lie_{\hat\xi_2}-\lie_{\hat\xi_2}\lie_{\hat\xi_1}=\lie_{[\hat\xi_1,\hat\xi_2]}$.
The necessary and sufficient condition for a vector field $\hat\xi^a\in\frakb$ is that there exists a function $\zeta:N\rightarrow\mathbb{R}$ such that
\begin{equation}\label{inbms}
  \lie_{\hat\xi}\hat h_{ab}=2\zeta \hat h_{ab},\quad \lie_{\hat\xi}\hat n^a=-\zeta \hat n^a.
\end{equation}

Despite the degeneracy of $\hat h_{ab}$, a covariant derivative can still be defined on $N$. For a covector field $\hat\mu_a$ with $\hat\mu_a\hat n^a=0$, the covariant derivative $D_a$ is defined in terms of the exterior derivative $\ud_a$ and the Lie derivative in the following way,
\begin{equation}\label{covdef}
  D_a\hat\mu_b=\frac{1}{2}(\ud_a\wedge\hat\mu_b+\lie_\mu \hat h_{ab}),
\end{equation}
where $\wedge$ donates the wedge product, and the vector $\hat \mu^a=\hat h^{ab}\mu_b$ in the subscript of the Lie derivative is ambiguous, but there is no ambiguity in the covariant derivative due to Eq.(\ref{bdg}). The generalization to $(0,l)$-type tensors is straightforward.

\subsection{Supertranslations}

Since $\hat n^a$ is a special vector field tangent to $N$, it is tempting to ask whether it is an infinitesimal symmetry. In general, this is not true. As a matter of fact, let $\lie_{\hat n}\hat h_{ab}=\frac{1}{2}\hat\theta\hat h_{ab}+\hat\sigma_{ab}$ where $\hat\theta$ is the expansion and $\hat\sigma_{ab}$ is the shear of the null geodesic congruence of the generators of $N$. Under the gauge transformation (\ref{conft}), it can be shown that,
\begin{equation}\label{cinhab}
  \lie_{\hat n'}\hat h'_{ab}=\omega(\lie_{\hat n}\hat h_{ab}+\hat h_{ab}\lie_{\hat n}\ln\omega),
\end{equation}
so
\begin{gather}
  \hat\theta'  =\omega^{-1}(\hat\theta+2\lie_{\hat n}\ln\omega),\label{eq1} \\
  \hat\sigma'_{ab} = \omega\hat\sigma_{ab},\label{eq2}
\end{gather}
Accordingly, if $\hat\sigma_{ab}=0$, a gauge transformation (\ref{conft}) with $\lie_{\hat n}\ln\omega=-\frac{1}{2}\hat\theta$ will make sure that $\lie_{\hat n'}\hat h'_{ab}=0$. In addition, $\lie_{\hat n'}\hat n'=0$, so $\hat n'$ is an infinitesimal transformation. A further gauge transformation with $\lie_{\hat n'}\omega'=0$ will preserve this result. But if $\hat\sigma_{ab}\ne0$, then $\lie_{\hat n'}h'_{ab}\ne0$, so $\hat n'$ is not an infinitesimal transformation. Therefore, for the null geodesic congruence tangent to $\hat n^a$, if it is \emph{shear-free}, there always exists a gauge such that under the flow of $\hat n^a$, the induced metric $\hat h_{ab}$ is Lie-dragged and $\hat n^a$ is an infinitesimal symmetry.

In particular, for the null infinity $\scri$ and the isolated horizon $\scrih$, a suitable gauge can be made such that $\hat n^a$ an infinitesimal symmetry. This gauge is called the \emph{Bondi gauge} and reads
\begin{equation}\label{bdg}
  \lie_{\hat n}\hat h_{ab}=0.
\end{equation}
This gauge fixing will be imposed in the following discussion. More generally, $\alpha\hat n^a$ is also an infinitesimal symmetry if and only if $\lie_{\hat n}\alpha=0$. These kind of infinitesimal symmetries are called the \emph{infinitesimal supertranslations}. For an infinitesimal supertranslation, $\alpha$ remains constant along the integral curves of $\hat n^a$, so does $\zeta$ for an arbitrary infinitesimal symmetry $\hat\xi^a$ in Eq.(\ref{inbms}). The infinitesimal supertranslations constitute a subalgebra $\frakt$ of $\frakb$. Since $\alpha$ is an arbitrary function, $\frakt$ is infinite dimensional. In addition, $\frakt$ is an Abelian subalgebra $\frakt$ of $\frakb$ because
\begin{equation}\label{stal}
  [\alpha\hat n,\beta\hat n]^a=(\alpha\lie_{\hat n}\beta-\beta\lie_{\hat n}\alpha)\hat n^a=0.
\end{equation}
Moreover, it is an ideal of $\frakb$ since for an arbitrary $\hat\xi^a\in\frakb$,
\begin{equation}\label{stidl}
  [\hat\xi,\alpha\hat n]^a=(\lie_{\hat\xi}\alpha-\alpha\zeta)\hat n^a,
\end{equation}
on the right side of which, the factor can be shown to be constant along the integral curves of $\hat n^a$, since $\lie_{\hat n}\lie_{\hat\xi}\alpha=(\lie_{\hat n}\lie_{\hat\xi}-\lie_{\hat \xi}\lie_{\hat n})\alpha=\lie_{[\hat n,\hat\xi]}\alpha=\lie_{\zeta\hat n}\alpha=0$.

\subsection{The infinitesimal Lorentz algebra}

Since $\frakt$ is an ideal of $\frakb$, the quotient $\frakl=\frakb/\frakt$ is well-defined and consists of equivalent classes of the infinitesimal symmetries whose elements differ from each other by some infinitesimal supertranslations.
The properties of elements of $\frakl$ can be understood using the covariant derivative defined on $N$.
 It can be easily checked that any $\hat\xi^a\in\frakb$ satisfies the following relations,
\begin{equation}\label{rels}
  \hat n^a\hat\xi_a=0,\quad D_{(a}\hat\xi_{b)}=\zeta\hat h_{ab},\quad \lie_{\hat n}\hat\xi_a=0,
\end{equation}
where $\hat\xi_a=\hat h_{ab}\hat\xi^b$. Conversely, any vector field $\hat\xi^a$ satisfying the above relations can be found to take the form $\hat\xi^a=\hat\eta^a+\alpha \hat n^a$, in which $\lie_{\hat n}\alpha=0$ and $\hat\eta^a$ also solves Eq.'s (\ref{rels}). In fact, adding an arbitrary infinitesimal supertranslation to $\hat\xi^a$ also gives rise to a solution to Eq.(\ref{rels}). Therefore, the solutions of Eq.(\ref{rels}) realize $\frakl$.

A natural fiber bundle structure of $N$ can be identified. The base manifold $S$ is a cross section of $N$ with $\hat n^a$ normal to it, and any integral curve of $\hat n^a$ intersects $S$ once and only once. The one-parameter diffeomorphism induced by $\hat n^a$ serves as the projection map $\pi$. The metric $\hat h_{ab}$ is the pullback of a suitable positive-definite metric $q_{ab}$ on $S$, i.e, $\hat h_{ab}=\pi^*q_{ab}$, and the covariant derivative $\hat D_a$ is also the pullback of the covariant derivative $\mathcal{D}_a$ on $S$ compatible with $q_{ab}$. The pushforwards of $\hat\xi^a$ and $\hat\xi'^a$ will be the same if $\hat\xi^a=\hat\xi'^a+\alpha\hat n^a$ for some function $\alpha$ with $\lie_{\hat n}\alpha=0$, and $\pi_*\hat\xi^a=\pi_*\hat\xi'^a=\varsigma^a$ are the solutions to the pushforward of Eq.(\ref{rels}), i.e.,
\begin{equation}\label{push}
  \mathcal D_{(a}\varsigma_{b)}=\zeta q_{ab},
\end{equation}
where $\zeta$ is viewed as a function defined on $S$ as $\lie_{\hat n}\zeta=0$. This implies that $\pi_*\hat\xi^a$ are the conformal Killing vector fields for $q_{ab}$. Since a 2-dimensional manifold equipped with an Euclidean metric is locally conformal to a flat space \cite{waldgr}, $\pi_*\hat\xi^a$ are generators of the conformal transformation on the 2-dimensional manifold, and so $\frakl$ is isomorphic to Lorentz algebra \cite{RN1116}. In particular, when $\scrin$ is $\scri$ or $\scrih$ (to be discussed below), the cross section $S$ has a topology of a 2-sphere, so $\frakl$ is isomorphic to the Lorentz algebra, too. Therefore, $\frakb$ can be viewed as an enhanced Poincar\'e algebra.


In the next two sections, the infinitesimal symmetries on the null infinity $\scri$ and the isolated horizon $\scrih$ will be identified and expressed in certain coordinate systems.
In the following discussion, the distinction between $\scrin$ and $N$ will not be spelled out. The restriction of a tensor field on $\scrin$ will be understood as the pullback to $N$ via $i^*$, if it can be pulled back.

\section{Asymptotic Symmetries on $\scri$}\label{secnull}

\subsection{The null infinity $\scri$}

The asymptotic symmetries exist on $\scri$ as a consequence of the asymptotic flatness of the physical spacetime $(M,g_{ab})$. Roughly speaking, a spacetime $(M,g_{ab})$ is said to be \emph{asymptotically flat} at the null infinity $\scri$ if there exists an embedding $\psi: M\rightarrow \tilde M$ from $M$ into the unphysical spacetime $(\tilde M,\tilde g_{ab})$ and $\tilde g_{ab}=\Omega^2 \psi_*g_{ab}$ with $\Omega$ a function on $\tilde M$ such that $\Omega=0$ and $\nabla_a\Omega\ne0$ on $\scri$. The physical spacetime is bounded by $\scri$ and the spatial infinity $i^0$. All null geodesics either come from the past null infinity $\scri^-$ or end on the future null infinity $\scri^+$, and all spacelike geodesics end at the spatial infinity $i^0$. $\tilde g_{ab}$ and $\Omega$ should be smooth on $\tilde M$ except at $i^0$. In addition, the topology of $\scri$ should be $S^2\times\mathbb R$, and certain causal conditions should be satisfied. More detailed definition of an asymptotic flatness on the null infinity can be found in \cite{waldgr,Ashtekar:2014zsa}.

By definition, the nonvanishing of the gradient of the conformal factor $\Omega$ on the null infinity $\scri$ suggests that the normal covector is naturally chosen to be $n_a=\tilde\nabla_a\Omega$. The smoothness condition imposed on $\tilde g_{ab}$ and $\Omega$, together with the vanishing of matter stress-energy tensor near $\scri$, implies that $n_a$ is null. It transforms properly according to
\begin{equation}\label{inatf}
  n'_a=\tilde\nabla_a(\omega\Omega)=\omega\tilde\nabla_a\Omega+\Omega\tilde\nabla_a\omega,
\end{equation}
under the gauge transformation Eq.(\ref{gag}). The last term vanishes on $\scri$, and thus the normal vector $\tilde n^a=\tilde g^{ab}n_b$ transforms according to $\tilde n'^a=\omega^{-1}\tilde n^a$. The induced metric $\tilde h_{ab}$ on $\scri$ of the unphysical $\tilde g_{ab}$ transforms as $\tilde h'_{ab}=\omega^2\tilde h_{ab}$. Therefore, there exists the conformally invariant quantity $\ghor=\tilde n^a\tilde n^b\tilde h_{cd}$. The Bondi gauge in this case is obtained by choosing a function $\omega$ such that
\begin{equation}\label{bdscri}
  \lie_n\ln\omega|_\scri=-\frac{1}{2}(\Omega^{-1}\tilde g^{ab}n_an_b)|_\scri,
\end{equation}
where all quantities on both sides are evaluated at the null infinity $\scri$. This leads to $(\Omega^{'-1}\tilde g^{'ab}n'_an'_b)|_\scri=0$, with $\Omega'=\omega\Omega$. In the following, the Bondi gauge is chosen, and primes will be dropped. The Bondi gauge is equivalent to $\lie_{\tilde n}\tilde g_{ab}|_\scri=0$ and $\tilde\nabla_a\tilde\nabla_b\Omega=0$. The second one implies that $n^b\tilde\nabla_bn^a=0$, i.e., the integral curves of $\tilde n^a$ are affinely parameterized geodesics on $\scri$, also called the generators of $\scri$ and they are also complete by the definition of the asymptotic flatness.

In the Bondi gauge, the Lie algebra $\frakb_\scri$ of infinitesimal symmetries includes $\alpha\tilde n^a$ with $\lie_{\tilde n}\alpha=0$ according to the general discussion in Section \ref{secgen}. $\alpha\tilde n^a$ are the infinitesimal supertranslations on $\scri$ and constitute the supertranslation Lie algebra $\frakt_\scri$ on $\scri$. Since the topology of $\scri$ is $S^2\times\mathbb R$, the quotient Lie algebra $\frakl_\scri=\frakb_\scri/\frakt_\scri$  is isomorphic to the Lie algebra of the conformal Killing vector fields on a 2-sphere, i.e., $\frakl_\scri$ is the Lie algebra of the infinitesimal Lorentz transformations on $\scri$.

\subsection{The infinitesimal symmetries on $\scri$}

The explicit form of an infinitesimal symmetry can be obtained by working in the Bondi-Sachs coordinates ($u,r,x^A$) with $A=2,3$. In this coordinate system, the metric takes the following form \cite{RN51},
\begin{equation}\label{bsmet}
\begin{split}
  \ud s^2=&\frac{V}{r}e^{2\beta}\ud u^2-2e^{2\beta}\ud u\ud r\\
  &+g_{AB}(\ud x^A-U^A\ud u)(\ud x^B-U^B\ud u),
\end{split}
\end{equation}
where $\beta, V, U^A, g_{AB}$ are functions of the coordinates, whose forms are to be determined by Einstein's vacuum equations with the following appropriate asymptotic behaviors,
 \begin{eqnarray}
 \beta=O(r^{-2}),\,V/r=O(r^{-1}),\,U^A=O(r^{-2}),\label{asym}
 \end{eqnarray}
and $g_{AB}$ is assumed to have the following asymptotic behavior,
\begin{equation}\label{gab}
  g_{AB}\ud x^A\ud x^B=r^2\gamma_{AB}\ud x^A\ud x^B+O(r),
\end{equation}
where $\gamma_{AB}$ can be taken to be the unit 2-sphere metric, i.e.,
\begin{equation}\label{gads}
  \gamma_{AB}=\left(
  \begin{array}{cc}
  0&\frac{2}{(1+z\bar z)^2}\\
    \frac{2}{(1+z\bar z)^2} & 0
  \end{array}\right),
\end{equation}
where $z=e^{i\phi}\cot\frac{\theta}{2}$ and $\bar z$ is the complex conjugate of $z$.

In this coordinate system, the null infinity $\scri$ is at $r\rightarrow\infty$. A conformal completion with $\Omega=1/r$ brings $\scri$ to a finite coordinate value $\Omega=0$, and the metric is
\begin{equation}\label{ngdd}
\begin{split}
  \ud\tilde s^2=&\Omega^2\ud s^2\\
  =&\Omega^3Ve^{2\beta}\ud u^2+2e^{2\beta}\ud u\ud\Omega\\
 & +(\gamma_{AB}+O(\Omega))(\ud x^A-U^A\ud u)(\ud x^B-U^B\ud u).
  \end{split}
\end{equation}
The normal covector to $\scri$ is $n_a=\tilde{\nabla}_a\Omega$, and
\begin{equation}\label{nU}
 \tilde n^a=e^{-2\beta}[(\pd_u)^a-\Omega^3V(\pd_\Omega)^a-U^A(\pd_A)^a],
\end{equation}
with $(\pd_x)^a=(\pd/\pd x)^a$.
Restricted to $\scri$,
\begin{equation}\label{noni}
  \tilde n^a = (\pd_u)^a.
\end{equation}
The restriction of $\tilde g_{ab}$ gives rise to the induced metric $\tilde h_{ab}$,
\begin{equation}
  \tilde h_{ab} = \gamma_{AB}(\ud x^A)_a(\ud x^B)_b, \label{inmeti}
\end{equation}
Here, the asymptotic behaviors Eq.(\ref{asym}) have been taken into account. This expression implies that the induced metric can be viewed as the round metric of a unit 2-sphere. It can be easily checked that the Bondi gauge is satisfied, as $\tilde g^{ab}n_an_b/\Omega|_\scri=0$.

Therefore, the supertranslations are given by the following vector fields on $\scri$,
\begin{equation}\label{sti}
  \tilde T^a=\alpha(z,\bar z)(\pd_u)^a,
\end{equation}
where $\alpha$ is an arbitrary function of variables $z,\bar z$. Now, let $\tilde\xi^a=\xi^u(\pd_u)^a+\xi^A(\pd_A)^a$ be the infinitesimal Lorentz transformation, so it satisfies Eq.(\ref{inbms}). The second relation implies that $\xi^u=u\zeta(z,\bar z)$ \footnote{The integration function $C(z,\bar z)$ is ignored since this has been considered in the supertranslation $\tilde T^a$.} and $\tilde n^a\pd_a\xi^A=0$, so $\xi^A=\xi^A(z,\bar z)$. The first relation determines $\zeta$ in terms of $\xi^A$,
\begin{equation}\label{zeta}
  \zeta=\frac{1}{2}\mathcal{D}_A\xi^A,
\end{equation}
where $\xi^A$ is viewed as a vector field tangent to the cross section $S$ of $\scri$. In addition, $\xi^z=\xi(z)$ and $\xi^{\bar z}=\bar\xi(\bar z)$. Therefore, the infinitesimal Lorentz symmetry is represented by,
\begin{equation}\label{ilt}
  \tilde\xi^a=\frac{u}{2}\mathcal{D}_A\xi^A\tilde n^a+\xi^A(\pd_A)^a.
\end{equation}
In summary, a general infinitesimal symmetry is thus given by,
\begin{equation}\label{gis}
  \tilde\xi^a=\left(\alpha+\frac{u}{2}\mathcal{D}_A\xi^A\right)\tilde n^a+\xi^A(\pd_A)^a,
\end{equation}
where $\alpha=\alpha(z,\bar z)$, $\xi^z=\xi(z)$ and $\xi^{\bar z}=\bar\xi(\bar z)$ are arbitrary functions.
The Lie bracket of two infinitesimal symmetries $\tilde\xi^a_1$ and $\tilde\xi^a_2$ is $\tilde\xi^a_3=[\tilde\xi_1,\tilde\xi_2]^a$, whose components are,
\begin{eqnarray}
  \xi^u_3 &=& \frac{1}{2}(\alpha_1\mathcal{D}_A\xi^A_2-\alpha_2\mathcal D_A\xi^A_1)+\xi^A_1\pd_A\alpha_2-\xi_2^A\pd_A\alpha_1
  \nonumber\\
  &&+\frac{u}{2}(\xi^B_1\mathcal D_B\mathcal{D}_A\xi^A_2-\xi^B_2\mathcal D_B\mathcal{D}_A\xi^A_1), \\
  \xi^A_3 &=& \xi^B_1\mathcal D_B\xi^A_2-\xi^B_2\mathcal D_B\xi^A_1.
\end{eqnarray}
Indeed, $\xi_3^A$ is a conformal Killing vector field for $\gamma_{AB}$, as its components are
\begin{eqnarray}
  \xi^z_3 &=& \xi_1\pd_z\xi_2-\xi_2\pd_z\xi_1,\label{eq-xiz3} \\
  \xi^{\bar z}_3 &=& \bar\xi_1\pd_{\bar{z}}\bar\xi_2-\bar\xi_2\pd_{\bar z}\bar\xi_1,\label{eq-xibz3}
\end{eqnarray}
which are functions of $z$ or $\bar z$, respectively.

Following \cite{RN51} and allowing $\alpha$ and $\xi^A$ to be holomorphic functions on the Riemann sphere, expand them in Laurent series as follows,
\begin{gather}
  \alpha_{n,l} =\frac{2}{z^n\bar z^l}, \\
  \xi_m = -z^{m+1},\quad\bar\xi_m=-\bar z^{m+1},
\end{gather}
where $l,n,m$ are all integers, the commutation relations become,
\begin{gather}
  [\xi_m,\xi_n]=(m-n)\xi_{m+n},\quad[\bar\xi_m,\bar\xi_n]=(m-n)\bar\xi_{m+n},
  \nonumber\\
  \quad[\xi_m,\bar\xi_n]=0, \label{eq-witt1}\\
  [\xi_m,\alpha_{n,l}]=\left(\frac{m+1}{2}-n\right)\alpha_{m+n,l},
  \nonumber\\
  [\bar\xi_m,\alpha_{n,l}]=\left(\frac{m+1}{2}-l\right)\alpha_{n,l+m}.
\end{gather}
Therefore, two copies of Witt algebra (\ref{eq-witt1}) is contained in the $\frakb_\scri$. Among these infinitesimal symmetries, $\alpha_{0,0},\,\alpha_{0,1},\,\alpha_{1,0},\,\alpha_{1,1}$ generate translations, while $\xi_{-1},\,\xi_0,\,\xi_1$ and $\bar\xi_{-1},\,\bar\xi_0,\,\bar\xi_1$ generate the global conformal transformations on the Riemann 2-sphere \cite{RN51}. These generators enable the definitions of the energy, momentum and angular momentum at the null infinity $\scri$.

In the usual approach, e.g., \cite{RN51}, the infinitesimal symmetries $\xi^a$ are obtained by requiring that,
\begin{gather}\label{barnich}
  \lie_\xi g_{rr}=\lie_\xi g_{rA}=g^{AB}\lie_\xi g_{AB}=0,
\\
  \lie_\xi g_{ur}=O(r^{-2}),\,\lie_\xi g_{uA}=O(1),
\\
  \lie_\xi g_{AB}=O(r),\,\lie_\xi g_{uu}=O(r^{-1}),
\end{gather}
that is, the transformation generated by $\xi^a$ preserves the asymptotic behavior of the metric $g_{ab}$.
Solving these relations gives the components of $\xi^a$ near $\scri$,
\begin{gather}
  \xi^u = T(z,\bar z)+\frac{1}{2}u\mathcal{D}_AY^A, \\
  \xi^r = -\frac{r}{2}\mathcal D_AY^A+\frac{1}{2}\mathcal D_A\mathcal D^A \xi^u+O(r^{-1}), \\
  \xi^A = Y^A-\frac{\gamma^{AB}\pd_B\xi^u}{r}+O(r^{-2}),
\end{gather}
where $Y^A$ is the conformal Killing vector field for $\gamma_{AB}$ with $Y^z=Y(z)$ and $Y^{\bar z}=\bar Y(\bar z)$. To obtain the infinitesimal symmetries, take the limit of $r\rightarrow\infty$ and drop the diverging $\xi^r$ to result in,
\begin{equation}\label{infsymbarn}
  \xi^a|_\scri=\left(T+\frac{u}{2}\mathcal{D}_AY^A\right)(\pd_u)^a+Y^A(\pd_A)^a,
\end{equation}
which agrees with Eq.(\ref{gis}). This result is expected because $\lie_{\tilde\xi}\ghor=0$ is equivalent to \cite{Geroch1977}
\begin{equation}\label{asymger}
  (\Omega^2\lie_\xi g_{ab})|_\scri=0.
\end{equation}
So the variation of $g_{ab}$ induced by the flow of $\xi^a$ is at most of the order $O(1/\Omega)$, i.e., $O(r)$.

\section{Asymptotic Symmetries on $\scrih$}\label{sechor}

\subsection{The isolated horizon $\scrih$}

A black hole in an asymptotic flat spacetime $(M,g_{ab})$ is a region from where no particle will arrive at the future null infinity $\scri^+$. The boundary of the black hole is called the event horizon, which is a global concept. This global nature narrows the application of the event horizon, e.g. it cannot be defined in a spatially compact spacetime. The concept of the event horizon is also teleological, which makes it less practical \cite{RN672}. So some quasi-local notations of black hole horizon have been proposed, and among them, \emph{isolated horizons} \cite{RN1054,RN1040,RN1042,RN1055,RN1056,RN1057,RN1058,RN1059,PhysRevD.67.024018} and \emph{dynamical horizons} \cite{RN1043,RN1061,RN1062,RN1063,RN1064,RN1065,RN1066,RN1067,RN1068,RN1069,RN1070,RN1071} are suitable for describing equilibrium and dynamical black holes.

In this work, the focus is mainly on the symmetries on the isolated horizon, as it is a null hypersurface, but the dynamical horizon is spacelike. In addition to being null, the isolated horizon is also a non-expanding horizon \footnote{Other conditions of the isolated horizon are unnecessary for the present discussion.}, that is,
\begin{equation}\label{nonexp}
  \lie_nh_{ab}=0,
\end{equation}
where $n^a$ is the normal vector field to $\scrih$ and $h_{ab}$ is the induced metric of the physical metric $g_{ab}$. This relation is equivalent to the statement that the area of the cross section $S$ on $\scrih$ does not change following the flow of $n^a$,
\begin{equation}\label{areaf}
  \lie_n\tilde\epsilon_{ab}=0,
\end{equation}
with $\tilde\epsilon_{ab}$ the induced volume element on $S$ satisfying $n^b\tilde\epsilon_{ab}=0$. This explains the nomenclature.

Under the conformal transformation Eq.(\ref{confdef}), the covector field $n_a$ is required to transform according to,
\begin{equation}\label{trna}
  \tilde n_a=\Omega n_a,
\end{equation}
which implies that $\tilde n^a=\Omega^{-1}n^b$. The induced metric $\tilde h_{ab}$ of the unphysical metric $\tilde g_{ab}$ which transforms according to $\tilde h'_{ab}=\Omega^2\tilde h_{ab}$, so there exists an gauge invariant structure $\ghor=\tilde n^a\tilde n^b\tilde h_{cd}$. Its symmetries can be called \emph{asymptotic horizon symmetries}, and the corresponding vector fields generating these symmetries are the \emph{infinitesimal horizon symmetries}.

In fact, the ``Bondi gauge" condition is already satisfied without further making a gauge transformation (\ref{conft}). However, it is still interesting to study the gauge transformations preserving the Bondi gauge. The conformal transformation Eq.(\ref{confdef}) violates the non-expanding condition Eq.(\ref{nonexp}), as
\begin{equation}\label{vionex}
  \lie_{\tilde n}\tilde h_{ab}=2h_{ab}\lie_n\Omega+\Omega\lie_nh_{ab}+2n_{(a}\nabla_{b)}\Omega.
\end{equation}
Although the last two terms vanish when restricted on $\scrih$, the first one does not in general. Now, choose a conformal factor such that $\lie_{n}\Omega=0$, and the Bondi gauge is recovered. So the following analysis will be done in the original gauge, i.e., $\Omega=1$.

There must exist infinitesimal horizon symmetries. Indeed, let $\xi^a=\alpha n^a$. It can be easily shown that $\lie_{\xi}  h_{ab}=0$. Therefore, in order for $\xi^a$ be an infinitesimal horizon symmetry, it is necessary that
\begin{equation}\label{reqal}
  \lie_{n}\alpha=0.
\end{equation}
It means that $\alpha$ remains constant along the integral curves of $n^a$ on the horizon $\scrih$. This is reminiscent of the supertranslation for $\scri$ when the Bondi gauge is chosen, so $\alpha n^a$ can be called the \emph{infinitesimal horizon supertranslation}.

The infinitesimal supertranslations $\alpha n^a$ constitute a Lie subalgebra named $\frakt_\scrih$, an Abelian ideal of $\frakb_\scrih$. By the definition of the isolated horizon $\scrih$, the cross section $S$ is diffeomorphic to a 2-sphere, so by the discussion in the last paragraph in Section \ref{secgen}, the quotient algebra $\frakl_\scrih=\frakb_\scrih/\frakt_\scrih$ is isomorphic to the 6-dimensional Lorentz algebra.

\subsection{The infinitesimal symmetries on $\scrih$}

In order to express the infinitesimal symmetry $\xi^a$ in an explicit form, a coordinate system need to be constructed. First, foliate the isolated horizon $\scrih$ by its cross sections $S_u$ where $u$ is a level function and constant on $S_u$, and
\begin{equation}\label{nvsv}
  n^a=(\pd_u)^a.
\end{equation}
Choose one of them, say $S_0$, which can be charted by an atlas of coordinate systems $x^A$ ($A=2,3$). The flow of $n^a$ will carry the coordinate systems of $S_0$ to each cross section $S_u$, so that locally, the horizon $\scrih$ is coordinatised by $(u,x^A)$. Second, let $l^a$ be the ingoing, future-pointing null vector field on $\scrih$ such that it is normal to $S_u$ and normalized according to $l_an^a=-1$. Smoothly extend $l^a$ off the horizon such that the vector field $l^a$ satisfies the geodesic equation, and is affinely parameterized by $\rho$, that is, $l^a=-(\pd/\pd\rho)^a$ and $\rho=0$ at $\scrih$. The flow of $l^a$ will map the coordinates ($u,x^A$) of $\scrih$ to a nearby hypersurface on which $\rho$ is constant. Therefore near the isolated horizon $\scrih$, the spacetime metric can be expressed in the ingoing Gaussian null coordinates ($u, \rho, x^A$) \cite{RN399},
\begin{equation}\label{nhmet}
\begin{split}
  \ud s^2=&(-2\ud u\ud\rho+q_{AB}\ud x^A\ud x^B)\\
  &+2\rho(\kappa\ud u^2+2\omega_A\ud v\ud x^A+k_{AB}\ud x^A\ud x^B)+\cdots.
\end{split}
\end{equation}
Here, $q_{AB}$ is the induced metric on the cross section $S_u$ on $\scrih$, $\omega_A$ is the connection 1-form on the normal cotangent bundle of $S_u$, $\kappa$ is the surface gravity for $n^a$ and $k_{AB}$ is the extrinsic curvature of $S_u$ for $l^a$. Finally, the ellipses stand for higher order terms in $\rho$.

The induced metric on the isolated horizon $\scrih$ is thus given by
\begin{equation}\label{inmet}
 h_{ab}=q_{AB}(\ud x^A)_a(\ud x^B)_b,
\end{equation}
and is Lie-dragged by the flow of $n^a$, i.e., Eq.(\ref{nonexp}), so
\begin{equation}\label{liedgh}
  \frac{\pd q_{AB}}{\pd u}=0,
\end{equation}
which means that $q_{AB}$ is not the function of $u$. Locally, $q_{AB}$ can always be put in the following form,
\begin{equation}\label{harm}
  q_{AB}=\Phi^2(x^C)\gamma_{AB},
\end{equation}
where $\Phi>0$, if a pair of conjugate harmonic function $x^C$ is chosen as the coordinates, and $\gamma_{AB}$ is given by Eq.(\ref{gads}).

Therefore, the infinitesimal horizon supertranslation is given by
\begin{equation}\label{horsuptrans}
 T^a=\alpha(z,\bar z)n^a=\alpha(z,\bar z)(\pd_u)^a.
\end{equation}
Let $\xi^a\in\frakl_\scrih$, then the second condition in Eq.(\ref{inbms}) implies that
\begin{equation}\label{horlorcond1}
  \xi^u=u\zeta,\quad\xi^A=\xi^A(z,\bar z).
\end{equation}
The first condition in Eq.(\ref{inbms}) determines the forms of $\zeta$ and $\xi^A$,
\begin{equation}\label{horlorcond2}
  \zeta=\frac{1}{2}\mathcal{D}_A\xi^A+\xi^A\pd_A\ln\Phi,\quad\xi^z=\xi(z),\quad\xi^{\bar z}=\bar\xi(\bar z),
\end{equation}
where $\mathcal D_A$ is the covariant derivative compatible with $\gamma_{AB}$.
Therefore, the infinitesimal Lorentz transformation is given by,
\begin{equation}\label{horlort}
  \xi^a=u\left(\frac{1}{2}\mathcal{D}_A\xi^A+\xi^A\pd_A\ln\Phi\right) n^a+\xi^A(\pd_A)^a.
\end{equation}

In summary, an arbitrary infinitesimal horizon symmetry is expressed as,
\begin{equation}\label{genhorinfsym}
  \xi^a=\left[\alpha+u\left(\frac{1}{2}\mathcal{D}_A\xi^A+\xi^A\pd_A\ln\Phi\right)\right] n^a+\xi^A(\pd_A)^a,
\end{equation}
where $\alpha=\alpha(z,\bar z)$, $\xi^z=\xi(z)$ and $\xi^{\bar{z}}=\bar\xi(\bar z)$. The Lie bracket of two infinitesimal symmetries $\xi^a_1$ and $\xi^a_2$ is $\xi^a_3=[\xi_1,\xi_2]^a$, whose components are,
\begin{eqnarray}
  \xi^u_3 &=& \frac{1}{2}(\alpha_1\mathcal{D}_A\xi^A_2-\alpha_2\mathcal D_A\xi^A_1)+\xi^A_1\pd_A\alpha_2-\xi_2^A\pd_A\alpha_1
  \nonumber\\
  &&+u\Big[\frac{1}{2}(\xi^B_1\mathcal D_B\mathcal{D}_A\xi^A_2-\xi^B_2\mathcal D_B\mathcal{D}_A\xi^A_1)
  \nonumber\\
  &&+(\xi^B_1\mathcal D_B\xi^A_2-\xi^B_1\mathcal D_B\xi^A_2)\mathcal D_A\ln\Phi\Big],\\
  \xi^A_3 &=& \xi^B_1\mathcal D_B\xi^A_2-\xi^B_2\mathcal D_B\xi^A_1.
\end{eqnarray}
$\xi_3^A$ is a conformal Killing vector field for $q_{AB}$, as its components are also given by Eq.'s (\ref{eq-xiz3}) and (\ref{eq-xibz3}).

If it is also permissible to expand $\alpha$ and $\xi^A$ in Laurent series on the Riemann sphere as follows,
\begin{gather}
  \alpha_{n,l} = \frac{2}{z^n\bar z^l}, \\
  \xi_m = -z^{m+1},\quad\bar\xi_m=-\bar z^{m+1},
\end{gather}
the commutation relations become,
\begin{gather}
  [\xi_m,\xi_n]=(m-n)\xi_{m+n},\quad[\bar\xi_m,\bar\xi_n]=(m-n)\bar\xi_{m+n},
  \nonumber\\
  \quad[\xi_m,\bar\xi_n]=0, \\
 [\xi_m,\alpha_{n,l}]=\left(\frac{m+1}{2}-n\right)\alpha_{m+n,l},
 \nonumber\\
 [\bar\xi_m,\alpha_{n,l}]=\left(\frac{m+1}{2}-l\right)\alpha_{n,l+m}.
\end{gather}
Therefore, two copies of Witt algebra are also contained in the $\frakb_\scrih$.
By analogue, $\alpha_{0,0},\,\alpha_{0,1},\,\alpha_{1,0},\,\alpha_{1,1}$ generate translations, while $\xi_{-1},\,\xi_0,\,\xi_1$ and $\bar\xi_{-1},\,\bar\xi_0,\,\bar\xi_1$ generate the global conformal transformations on the Riemann 2-sphere. These generators also make it possible to define of the energy, momentum and angular momentum at the isolated horizon $\scrih$.

\section{Conclusion}

In this work, the asymptotic symmetry of a null hypersurface $\scrin$ is investigated based on the method inspired by the geometric treatment of Geroch \cite{Geroch1977}. The key concept in Geroch's treatment is the conformally invariant tensor $\ghor$ which fully describes the intrinsic geometry of the null infinity $\scri$ by incorporating its normal vector field and the induced metric. The infinitesimal symmetry is thus defined as the vector field tangent to the null infinity $\scri$ generating the conformal transformation. This method has been easily extended to other null hypersurfaces, in particular, the isolated horizon $\Delta$. The infinitesimal symmetries form an infinite dimensional Lie algebra $\frakb$. It contains an Abelian ideal, which is called the supertranslation Lie algebra $\frakt$. It consists of tangent vector fields parallel to the generators of $\scrin$, whose flows preserves the induced metric as well as the generators. The quotient algebra is the asymptotic Lorentz algebra $\frakl$, which is isomorphic to the Lorentz algebra, just as its name implies. This enlarges the Poincar\'e algebra. As special examples, the asymptotic symmetries of the null infinity $\scri$ and the isolated horizon $\scrih$ are discussed. This approach yields the similar asymptotic symmetry generators as obtained using perturbative methods, previously.

\begin{acknowledgements}
The author wishes to thank Yungui Gong for his help. This research was supported in part by the Major Program of the National Natural Science Foundation of China under Grant No. 11475065 and the National Natural Science Foundation of China under Grant No. 11690021.
\end{acknowledgements}

\bibliographystyle{apsrev4-1}
\bibliography{../../References/BMSHorizon}

\end{document}